# Resilience Dynamics in Coupled Natural-Industrial Systems: A Surrogate Modeling Approach for Assessing Climate Change Impacts on Industrial Ecosystems

*William Farlessyost[1] and Shweta Singh[1,2,3,*]*
[1]*Agricultural & Biological Engineering, Purdue University*
[2]*Environmental & Ecological Engineering, Purdue University*
[3]*Davidson School of Chemical Engineering, Purdue University*
[*]*Corresponding Author*

# Abstract

*Industrial ecosystems are coupled with natural systems through utilization of feedstocks and waste disposal. To ensure resilience in production of industrial systems under the threat of climate change scenarios, it is necessary to evaluate the impact of this coupling on productivity and waste generation. In this work, we present a novel methodology for modeling and assessing the resilience of coupled natural-industrial ecosystems under climate change scenarios. We develop a computationally efficient framework that integrates liquid time-constant (LTC) neural networks as surrogate models to capture complex, nonlinear dynamics of coupled agricultural and industrial systems. The approach is demonstrated through a case study of a soybean-based biodiesel production network in Champaign County, Illinois. LTC models are trained to capture dynamics of nodes and are then coupled and driven by statistically downscaled climate projections for RCP 4.5 and 8.5 scenarios from 2006-2096. The framework enables rapid simulation of system-wide material flow dynamics and exploration of cascading effects from climate-induced disruptions. Results reveal non-linear behaviors and potential tipping points in system resilience under different climate scenarios and farm sizes. The RCP 8.5 scenario led to earlier and more frequent production failures, increased reliance on imports for smaller farms, and complex patterns of waste accumulation and stock levels. The methodology provides valuable insights into system vulnerabilities and adaptive capacities, offering decision support for enhancing the resilience and sustainability of coupled natural-industrial ecosystems in the face of climate change. The framework's adaptability suggests potential applications across various industrial ecosystems and climate-sensitive sectors.*

# Introduction

Climate change poses significant challenges to global food security and sustainable energy production, necessitating a deeper understanding of coupled natural-industrial systems' resilience [1], [2]. The increasing frequency and severity of extreme weather events, shifts in precipitation patterns, and rising temperatures are altering the fundamental dynamics of agricultural ecosystems [3], [4], [5], [6], [7], [8], [9], [10], [11]. These changes have far-reaching implications for crop yields, water availability, and the overall stability of food production systems. As climate variability intensifies, it becomes crucial to develop robust methodologies for assessing and predicting the impacts on interconnected natural and industrial ecosystems, particularly in the context of agricultural value chains and their associated energy production systems.

Industrial ecosystems, particularly those centered around agricultural production and processing, are inextricably linked with natural systems through complex feedbacks and dependencies [2],[12], [13], [14], [15]. These coupled natural-industrial systems exhibit highly intricate and nonlinear dynamics driven by diverse interacting factors ranging from resource constraints to market forces [1], [2]. For instance, in the case of biofuel production, the industrial ecosystem encompasses not only the manufacturing processes but also the agricultural systems that supply the raw materials. This coupling manifests through various pathways, including land use changes, water consumption, nutrient cycling, and energy flows. The interdependencies between crop growth, harvesting, processing, and end-product distribution create a network of interactions that span multiple spatial and temporal scales, making these systems particularly vulnerable to climate-induced perturbations.

The cascading effects of climate change-induced disruptions in natural systems can significantly impact the stability and efficiency of industrial ecosystems [16]. As climate variability alters crop growth patterns, yield potentials, and water availability, these changes propagate through the entire value chain, affecting industrial processes, resource allocation, and ultimately, product output [17], [18]. For example, shifts in temperature and precipitation regimes can lead to changes in crop phenology and productivity, which in turn influence the timing and volume of raw material inputs to processing facilities [14]. This can result in supply chain disruptions, necessitating adaptive measures in industrial operations. Moreover, extreme weather events can cause acute shocks to the system, potentially leading to crop failures or infrastructure damage that ripple through the interconnected network of agricultural and industrial processes. Understanding these cascade effects is crucial for developing resilient and sustainable industrial ecosystems in the face of ongoing climate change.

To analyze the intricate relationships within coupled natural-industrial ecosystems, researchers have traditionally employed various static methodologies. Life Cycle Assessment (LCA) and Material Flow Analysis (MFA) have been widely used to quantify the environmental impacts and resource flows within these systems [19], [20], [21], [22]. These approaches provide valuable insights into the overall sustainability and efficiency of industrial processes, including their interactions with natural ecosystems. For instance, LCA can help evaluate the environmental footprint of agricultural products throughout their lifecycle, from crop cultivation to final consumption. Similarly, MFA allows for tracking material and energy flows across the industrial ecosystem, revealing potential inefficiencies and connections with environmental systems. While these static methods offer a comprehensive snapshot of system performance, they often struggle to fully capture the dynamic nature of climate-induced disruptions and the resulting cascade effects on coupled natural-industrial systems [23], [24], [25].

The complex interactions within coupled natural-industrial systems necessitate accounting for intricate multiscale dynamics. These systems exhibit behaviors that span various temporal and spatial scales, from rapid biochemical reactions in crop growth to long-term climate trends affecting agricultural productivity. Capturing these multiscale dynamics is crucial for understanding system resilience and predicting responses to climate change. For example, short-term weather fluctuations can impact daily crop water requirements, while long-term climate shifts may alter entire growing seasons and crop suitability in a region. Similarly, industrial processes may have immediate responses to input variations, as well as long-term adaptations to persistent changes in raw material quality or quantity. Integrating these diverse scales of interaction is essential for a comprehensive understanding of system behavior under different climate change scenarios [26], [27]. This multiscale approach allows for bridging the gap between natural and industrial ecosystem modeling, providing a more nuanced view of potential vulnerabilities and adaptive capacities within the coupled system.

While researchers currently employ various methods such as system dynamics, agent-based models, and mechanistic crop simulations to capture dynamic behaviors, these traditional system identification and modeling techniques often struggle to fully capture the intricate physical and informational interdependencies underlying real-world industrial systems. System dynamics approaches, while adept at representing material, information, and cash flows between entities [4], [19], [20] often fall short in accurately capturing the intrinsic physical and chemical transformations within production processes [23], [24], [25], [26], [27]. Dynamic material flow analysis (DMFA) offers insights into temporal changes in material stocks and flows but can be limited in capturing complex feedback mechanisms and non-linear behaviors in coupled systems [28], [29], [30], [31]unless mechanistic details of processes are included in projection. Agent-based models offer flexibility in mapping behaviors and interactions but can grow exponentially complex [32], [33], [32], [34], [35], [36], [37], [38]. Mechanistic models of crop growth, nutrient cycling, and biogeochemistry [39], [40], [41], [42], [43], [26] while detailed, are often constrained by uncertainties in model parameters and future conditions, limiting their utility for long-term scenario planning. Moreover, these approaches can be computationally intractable for long-term simulation and study of large-scale coupled systems. As a result, there is a critical need for modeling approaches that can effectively couple natural and industrial ecosystems, capturing their complex interactions while maintaining computational feasibility.

To address this critical gap and provide actionable insights for decision-makers, our study presents a broad framework for modeling these coupled dynamics using a system identification-based approach. This novel methodology aims to capture the complex, nonlinear dynamics across entire industrial ecosystems, offering a more comprehensive understanding of system behavior under different climate change scenarios. By employing data-driven techniques, we seek to overcome the limitations of traditional modeling approaches and bridge the gap for modeling dynamics of coupled natural and industrial ecosystems. The proposed approach utilizes a system identification approach of key system components and then modeling their interactions, facilitating a more accurate representation of the coupled natural-industrial ecosystems dynamics.

To implement this system identification-based approach, we leverage surrogate modeling techniques, specifically using data-driven liquid time constant (LTC) neural networks [44]. Surrogate models provide computationally efficient approximations of complex underlying processes, allowing for rapid scenario analysis and system-wide simulations that would be impractical with full mechanistic models. LTC neural networks offer unique advantages in capturing temporal dynamics and nonlinear behaviors inherent in coupled natural-industrial ecosystems [45], [46]. These networks employ 'liquid' neurons with dynamic time constants, enabling the capture of complex temporal contexts and nonlinear dynamics without the computational burden of full mechanistic simulations. By training these LTC models on synthetic data generated from high-fidelity simulations, we combine the detail of mechanistic models with the efficiency of more abstract representations [47]. This novel combination provides crucial supervision and regularization, allowing the model to converge on meaningful causal relationships while avoiding the parameter uncertainties and computational limitations often encountered in fully mechanistic approaches [46], [48].

To demonstrate the proposed methodology, we present a case study focusing on a soybean-based biodiesel industrial ecosystem in Champaign County, Illinois. This coupled natural-industrial ecosystem encompasses the entire value chain, from soybean agriculture to biodiesel production. The industrial ecosystem in this case consists of manufacturing nodes representing soybean oil extraction and biodiesel production processes. We apply the LTC modeling approach to create surrogate models that efficiently simulate the growth patterns of soybean crops, soybean oil production, and conversion to soybean biodiesel. These surrogate models are driven by statistically downscaled climate projections from 2006-2096 under two scenarios: Representative Concentration Pathway (RCP) 4.5 and 8.5 [49]. This setup enables us to analyze potential future climate impacts on crop yields and subsequent cascading effects throughout the integrated system with a level of computational efficiency that allows for extensive exploration of scenarios and sensitivities.

The remainder of this paper is structured in the following way. ***Methods*** section describes our methodology in detail, outlining the steps for system boundary definition, component system identification, data collection and preprocessing, model architecture and training, and the coupling of dynamics across the natural-industrial ecosystem. We also discuss the development of algorithmic controllers and the generation of exogenous climate and economic demand signals. ***Results*** section presents the case study setup, detailing the specific components of the soybean biodiesel industrial ecosystem in Champaign County, Illinois and provides a comprehensive analysis of our results, discussing the simulation outputs that assess production rates, waste generation, stock levels, and required imports under RCP 4.5 and 8.5 scenarios for varying farm sizes. We evaluate the system's resilience and sustainability under different climate projections and discuss the implications of our findings. Finally, ***Conclusions and Discussions*** section concludes the paper by summarizing key insights, addressing limitations, and suggesting directions for future research in modeling coupled natural-industrial ecosystems under climate change.

# Methods

The proposed approach for modeling coupled natural-industrial network dynamics and using these models to study resilience dynamics to maintain production involves three distinct parts: 1) Defining system and modeling the nodes 2) Coupling Dynamic models and 3) Algorithm Controller Design & Exogenous Drivers Modeling. Figure 1 shows the flow diagram of interactions between these steps. Each parts involve several steps, that we describe in detail here.

### Step 1: System Definition & Node Modeling

**System Boundary**: We start by defining a system boundary to demarcate the spatial regions of interest, ensuring that all industrial operations and natural ecosystem dynamic processes and interactions of interest are encapsulated while extraneous influences are excluded. Industrial operations usually entail various production processes, and thus the system boundary should delineate the specific life cycle stages, from raw material sourcing to the delivery of end products, being considered. Natural ecosystems reflect the dynamics of relevant crops or plant growth, considering environmental variables that influence growth patterns and serve as the principle dynamic drivers. Dynamic processes and interactions from either natural or industrial systems for which data does not exist to build state-accurate models should be excluded from the system boundary, however, their impact can be considered as an exogenous perturbation to the system. Care should be taken to ensure the system boundary remains realistic over whatever time intervals are considered as the period of analysis[50].

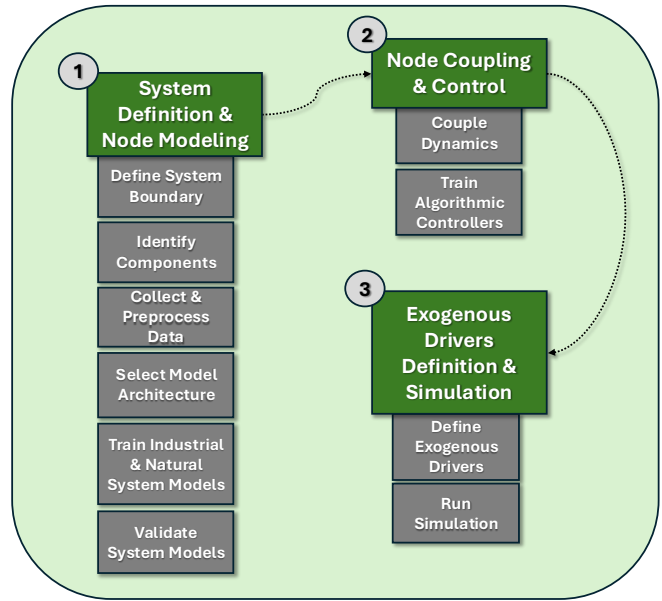

*Figure 1 : Methodology for Modeling Dynamics of Coupled Natural-Industrial Ecosystems*

**Component Identification**: Identifying and understanding the individual components within the system boundary is paramount for a thorough analysis. This involves breaking down the system into discrete nodes or units, each representing a specific process or entity with clearly defined inputs and outputs, either from or to another node, or from or to an exogenous source/sink. For instance, in an industrial setting, nodes might represent distinct production stages or units with feedstock and material flowrates, while in a natural ecosystem, nodes could signify the growth of species or natural material availability over time given changes in environmental conditions exogenous to the system or anthropomorphic waste flows endogenous.

For each identified node, it is essential to define its characteristics. This should involve detailing its operational parameters in an industrial context or growth patterns and interactions in an ecological one. Failure to understand the operating ranges, scales, and frequency of this system behavior prior to developing and connecting node models, will result in physically unrealistic simulated system state trajectories at best, but more likely integrated coupled-model failure.

**Data Collection and Preprocessing** Further, understanding the characteristics of each node allows us to choose the appropriate data on which to train individual node models. The emphasis throughout this process is to ensure that the data is both representative of the real-world dynamics of the node and is structured in a way that is conducive to effective modeling. Proper preprocessing not only enhances the accuracy of the models but also contributes to their generalizability across different scenarios or conditions. Data can be sourced from a combination of experimental datasets, historical records, and synthetic data generated from mechanistic models. Depending on the specific node in question, this might involve time-series data capturing dynamic changes, cross-sectional data detailing model parameterization in specific instances, or even qualitative data that is subsequently quantified for modeling purposes to scale the data/model.

Given the diverse sources of data and the potential variations in scales, units, or magnitudes, standardizing or normalizing the data prior to use in training becomes imperative. This ensures that no particular feature or variable disproportionately influences the model. Techniques like z-score normalization or Min-Max scaling are employed based on the data distribution. Additionally, in datasets with high noise, smoothing techniques, such as rolling averages, might be applied where appropriate to focus on underlying trends or patterns.

**Model Architecture and Training** Building an accurate model for node dynamics necessitates a careful selection of model architecture and a structured training process that may vary from node to node. The following outlines the general approach to model architecture and training.

A suitable model architecture should be chosen depending on the nature and characteristics of the data. The primary criterion for model selection is its ability to capture the underlying patterns and dynamics of the node in question and predict the dynamics of the modeled outputs given a set of input dynamics. Second to this, is interpretability/auditability of the model structure and parameterization. In general, the following modeling approaches may be suitable:
   a) Low-order, first principle, ODE models well established against real-world data,
   b) Recurrent neural networks (RNNs) [51] or their variants such as Long Short-Term Memory (LSTM) [52] and LTC neural networks [44],
   c) SINDy (Sparse Identification of Nonlinear Dynamics) or Symbolic Regression derived ODE models [53], [54], [55], [56].

Available datasets for each node are typically partitioned into training, validation, and testing subsets. The training set is used to adjust the model's parameters, the validation set aids in hyperparameter tuning and to prevent overfitting during training, and the testing set provides an unbiased performance evaluation. Training is typically iterative, where the model is exposed to the data multiple times (epochs), refining its weights and biases to minimize the discrepancy between its predictions and actual values.

**Model Evaluation** The evaluation process not only gauges the model's performance on unseen data but also informs potential refinements or improvements so that the developed model is robust, accurate, and generalizable. Each node model should be integrated or computed over time and the output should be compared against the ground truth data trajectory.

A summary quantitative metric such as Mean Squared Error (MSE), Root Mean Squared Error (RMSE) or Mean Absolute Error (MAE) should be employed. These metrics provide a measure of the model's ability to make accurate predictions that can be easily compared. However, beyond quantitative metrics, visual inspections play a crucial role in understanding the model's performance. By plotting actual vs. predicted values, inconsistencies, outliers, or patterns of discrepancies can be visually identified. These plots offer an intuitive understanding of where the model excels and where it may falter and should be improved. This is especially useful in scenarios where quantitative metrics might not capture nuanced discrepancies or where the nature of errors is as important as their magnitude.

## Step 2: Node Coupling & Control

It is crucial that all interaction between the industrial and natural components identified as nodes within the system boundary is appropriately characterized through node-to-node coupling of dynamics. This ensures that the dynamic responses are accurately propagated throughout the system. Each node's output is identified and mapped to serve as the direct input for a subsequent node or nodes, or an exogenous sink outside the system boundary. Likewise, each node input is either mapped as the output of another node or an exogenous source. Consequently, it is vital that all node models and exogenous data sources are on the same timescale. To this end, retiming or interpolation of either a model or data may be necessary to achieve this.

Some nodes may direct output to stock variables, which serve as reservoirs or accumulative entities. These stock variables should be identified, and their dynamics should be defined with the appropriate difference equations (i.e. the input and output relationships affecting accumulation, and rate of deterioration and transfer to waste). Stock variables can, in turn, influence other nodes, either by serving as direct inputs or by affecting the conditions governing a node's behavior.

In certain scenarios, the transmission of an output from node A to the input of node B might be heavily delayed or have significant consequences in the overall system dynamics. Such delayed responses should be accounted for, ensuring that the system's dynamics capture these temporal lags. Methods range from adding a simple time-delayed lag into a signal's time-series feed, to building an additional dynamic model to approximate this lag from known data.

## Step 3: Exogenous Drivers Definition & Simulation

**Algorithmic Controllers in Node Integration**: Within the context of the simulated system, we suggest that "controllers" should be employed as algorithmic constructs that ensure the right balance between demand and the necessary feedstock for industrial nodes. By adjusting inputs based on real-time simulated demand, the system can emulate the adaptive nature of real-world industrial processes seeking to produce enough to hit that changing demand.

Therefore, the controllers should continuously monitor the simulated economic production demand and based on the current demand and setpoint values, the algorithmic controllers fine-tune the feedstock quantities. This ensures that the simulated industries operate in line with their demand profiles, even in scenarios where there are fluctuations or changes in the demand. Outputs from industrial nodes, regulated by these algorithmic controllers, serve as inputs to subsequent nodes or systems in the simulation. By embedding these algorithmic controllers within the simulation, the methodology ensures that each node or industrial process is responsive and adaptive, reflecting the dynamic interplay between demand and supply in a realistic manner.

**Simulation of Exogenous Disruptors and Economic Demand Scenarios** Understanding how the coupled natural-industrial ecosystems respond to external scenarios, especially those that vary depending upon the placement and drawing of the system boundary spatially, is necessary for gauging its resilience and overall sustainability. To this end, we suggest supplying two types of exogenous time-series inputs to the system: *disruptive signals* and *economic demand signals*. "Disruptive signals" propagate the dynamics of a process or event that might have disruptive consequences on a system node (e.g. drought, flooding, infrastructure damage, etc.) through the system via the information as node input. "Economic demand signals" function as the production goal of the network over time. These signals are passed to the algorithmic controllers, which translate them into feedstock requirements for each of the industrial node models to meet demand. Depending on the node or component, these external forces might directly influence production rates, resource availability, or even operational feasibility.

Depending on the focus of the study, the simulation may be run over varying time horizons. Short-term simulations might capture immediate system responses, while long-term simulations can provide insights into cumulative effects and potential tipping points.

**<u>Evaluating System Resilience</u>**: We assess the system's resilience, here defined as its ability to maintain production matching the economic demand signal given the exogenous disruptor signal, using the generalized approach given in Figure 2.

# Results

We use our proposed methodology to evaluate the resilience of a coupled natural-industrial ecosystem shown in Figure 3. In this network, industrial systems of soybean oil and soybean biodiesel is coupled with the natural system of soybean farming, thus forming a coupled natural industrial ecosystem. Further, we test the resilience of this network to climate change disruptions under RCP 4.5 and 8.6 for meeting the simulated economic demand scenarios. We show the system boundary, results on training surrogate model for node dynamics and coupled dynamics under climate change impact here. The details of data generation, preprocessing and economic and climate forecasts are provided as the first and second SI sections.

1. ## System Definition & Node Modeling

**<u>System Boundary & Component Identification</u>**

```
// 1. Define minimum and maximum acceptable thresholds for each stock variable or reservoir
FOR EACH stock_variable IN system
    DEFINE min_threshold, max_threshold
    // These thresholds represent levels below which production might be compromised
    // or above which there might be wastage or inefficiencies
END FOR
// 2. Track stock level fluctuations in response to the disruptor signal
FOR EACH stock_variable IN system
    MONITOR stock_level
    IF stock_level DEVIATES significantly FROM min_threshold OR max_threshold THEN
        IDENTIFY system vulnerabilities OR inefficiencies
    END IF
END FOR
// 3. Evaluate stock variables as buffers against sudden changes in demand or supply
FOR EACH stock_variable IN system
    ASSESS buffer_capacity
    IF stock_variable ABSORBS disruptions WITHOUT affecting production THEN
        CLASSIFY stock_variable AS robust
    ELSE
        NOTE potential for wastage OR inefficiencies
    END IF
END FOR
// 4. Track production rates of nodes in the network throughout the simulation
FOR EACH node IN network
    TRACK production_rate DURING simulation
    IF production_rate IS consistent EVEN in adverse scenarios THEN
        INDICATE strong system resilience
    ELSE
        FLAG potential weaknesses
    END IF
END FOR
// 5. Measure quantities and frequencies of imports required to maintain stability and production levels
TRACK import_quantities, import_frequencies
IF imports ARE frequent OR large THEN
    INDICATE system vulnerabilities OR deficiencies
END IF
// 6. Identify tipping points where the system might bifurcate from a stable state to failure
FOR EACH stock_variable IN system
    FOR EACH node IN network
        IF stock_levels OR production_rates REACH critical levels THEN
            IDENTIFY tipping_point
        END IF
    END FOR
END FOR
// 7. Evaluate waste generation trends within the system
TRACK waste_generation
FOR EACH stock_variable IN system
    IF unused_stock IS generated AND exceeds acceptable levels THEN
        IDENTIFY inefficiencies OR overestimations in stock procurement
    END IF
END FOR
```

*Figure 2: System resilience assessment methods.*

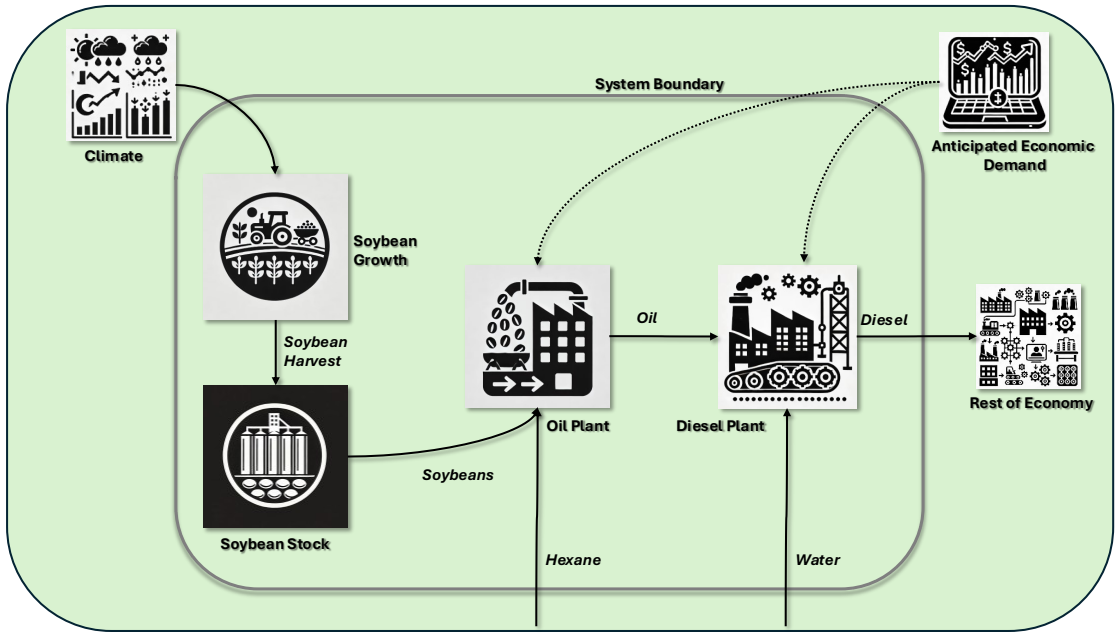

*Figure 3: Hypothetical soybean network in Champaign, Illinois.*

Spatially, we draw our system boundary to encapsulate the area directly surrounding Champaign, Illinois. The system boundary includes soybean agricultural production as well as derivative products of oil and diesel. The soybean production network works to meet the external economic demand for diesel, while also impacted by the exogenous disruptor of changing climate conditions, affecting the growth of soybeans.

At the heart of the network lie three dynamic nodes, as shown in Figure 3. These nodes represent distinct life cycle stages in the soybean biodiesel production.

*Soybean growth:* this node models the seasonal growth of soybeans, heavily influenced by various climate variables such as temperature, precipitation, and solar radiation. The culmination of this stage is the harvest, with the yield being transferred to an inventory stock. We consider soybean farms of 450, 500, and 550 hectares in size.

*Oil plant:* acting as a bridge between raw soybeans and their derivative products, this node employs solvent extraction using hexane to convert soybeans from the inventory into crude soybean oil. An important by-product of this stage is soybean meal.

*Diesel plant:* the culmination of the value chain, this node takes the crude soybean oil and, in conjunction with water, processes it into diesel fuel via trans-esterification. This diesel is passed to some exogenous demand sink outside of the system boundary.

A *soybean stock* is employed as dynamic entity within this network defined as the difference between yearly soybean harvest added and soybean feedstock utilized for oil. It also acts as a buffer, whose level determines whether the soybean oil production node has a consistent supply, ultimately leading to a steady production of soybean diesel. However, stock that accumulates and is not used within a three-year period is removed and sent to *waste*.

We make several assumptions about network operation to simplify the analysis. Firstly, soybean growth is assumed to be homogenous across the farm and the harvest time does not change year to year. All harvested soybeans are accumulated in the soybean stock, and the soybean plant is the only user of this stock. Further, all soybean oil is used in the production of soybean diesel and all soybean diesel produced is used to satisfy exogenous economic demand. Other necessary feedstocks that vary dynamically over time (i.e. water and hexane) are considered as exogenous sources, whose utilization is not tracked.

**Node LTC Model Architecture**

Our modeling approach for the industrial and natural dynamic nodes utilizes LTC neural networks. Three different types of neurons make up each model. Input neurons in an LTC network are designed to interface with the environment, receiving external time-varying signals. Unlike traditional artificial neurons, LTC input neurons can handle continuous-time inputs, which allows them to process information that changes over time, such as the varying rates of soybean and hexane input in a soybean oil plant. Hidden neurons in LTC networks are critical for capturing the temporal dynamics of the input signals. Each hidden neuron has a liquid time-constant that governs its rate of change, allowing the neuron to adapt its response over time. This is particularly important for modeling industrial processes like those in soybean oil and biodiesel plants, where the state of the system evolves continuously, and the reaction times of different processes vary. The hidden neurons integrate information over time, learning not just from the current input, but also from the history of received inputs, which is essential for predicting and understanding complex, time-dependent phenomena. Output neurons, or motor neurons, in an LTC network generate the final output signals that control actions or make predictions based on the processed inputs and the learned temporal dynamics [44]. In the context of the soybean oil and biodiesel plant models, these neurons output signals correspond to the production rates.

*Industrial System Node Models*

For the soybean oil plant, the input nodes consist of soybean and hexane, which are essential to the operation of the plant (Figure 3). In the case of the biodiesel plant, the inputs are water and soybean oil, marking the beginning of the biodiesel production process. The hidden nodes, which serve as the intermediary layer in both models, are determined through an iterative process. The number of hidden neurons is increased until the integrated model no longer approximated a horizontal line. Finally, the motor neurons are selected based on their value-added significance and the necessity for a proper model fit. For the soybean oil plant, the outputs are soybean oil and soymeal, while for the biodiesel plant, the outputs are diesel and an oil recycling stream internal to the plant.

*Natural System Node Model*

For the modeling of soybean growth, the LTC neural network is configured with an emphasis on key environmental variables. The inputs to this model are streamlined to the most impactful factors: time, precipitation, and temperature. These sensory neurons were selected through an iterative process seeking to reduce the number of model inputs without wrecking model performance, while holding the number of hidden neurons constant (31 hidden neurons with 1 motor neuron). With inputs determined, the hidden layer, composed of 20 neurons, was determined through an iterative approach akin to that utilized for industrial plant modeling. At the output layer, a single motor neuron corresponds to the overall soybean yield. This output reflects the model's targeted purpose, to accurately represent the biological growth cycle of soybean plants under varying environmental conditions.

**LTC Model Training**

The four LTC models are trained on the synthetic data discussed previously with an 80-20 training-to-testing split. The soybean oil and diesel production plant models were thus each trained across 8,000 hours of data, while the soybean growth, with different models for each RCP 4.5 and 8.5, were trained across 72 growing seasons (also at hourly resolution).

The training loss for the soybean oil LTC model, shown in SI Figure 4, exhibited a rapid decline during the initial stages. Convergence was indicated as the loss plateaued around 125 epochs. To mitigate the risk of overfitting due to inherent noise within the training data, training was curtailed at this juncture.

Akin to the soybean oil production model, the biodiesel production LTC model's training loss saw a swift reduction, stabilizing around the 125-epoch mark. Training was subsequently terminated to avert potential overfitting.

Distinctively, the soybean growth LTC model's training loss experienced a sharp decrease post the inaugural pass through the multi-year seasonal dataset. The loss consistently remained below a 0.01 mean squared error in the ensuing cycles, displaying a discernible repetition around the 400th step. Given the overarching goal to thwart potential overfitting, training was concluded at 200 steps.

**LTC Model Evaluation**

The LTC node models were rigorously evaluated against 20% of the generated synthetic data to gauge their proficiency in capturing the intricate dynamics of the soybean production system. For the soybean growth model, this equates to 18 growing seasons. The models are integrated and compared against this test data in SI Figure 5.

The soybean diesel model displays good fit with a root mean squared error (RMSE) of 0.11 on the test data. Noteworthy was its capability to accurately pinpoint key inflection points and shifts within the diesel production time series. However, achieving the absolute correct value of local extrema remained elusive to the model's predictions. Achieving a test RMSE of 0.194, the soybean oil LTC model's performance was congruent with the diesel model. Its adeptness at capturing trend reversals was excellent, manifesting its ability to mirror the actual soybean oil production trajectory.

The soybean growth model rendered an RMSE of 0.105 on the test set. The model's predictions did exhibit sporadic oscillations, particularly before the onset of full germination and during intervals characterized by swift growth. Nonetheless, its predictions closely aligned with the long-term mass trajectory, effectively identifying pivotal moments of crop maturity and peak harvest mass.

## 2. Node Coupling & Control

**Algorithmic Controller Model Architecture, Training Method & Evaluation**

We use LTC neural networks here as our algorithmic controllers for prediction of node model feedstocks required to meet economic demand. A corresponding algorithmic controller is trained for both the soybean oil and soybean diesel plants. The design is such that the diesel plant controller takes in a demand timeseries for diesel (the network production goal) and estimates the oil and water feedstocks required by the diesel plant to meet this goal. The desired oil feedstock is passed to the oil plant controller, which again predicts the full timeseries of necessary soybean and hexane feedstocks for the oil plant model. The oil plant, taking both the predicted hexane and soybean feedstocks as inputs, gives the actual oil produced as an output, which is further fed into the diesel plant along with the predicted required water feedstock.

The algorithmic controller LTC models are trained on top of the previously learned LTC industrial plant model. The desired production stream time series is provided to the LTC controller, and the prediction is then passed through the LTC industrial models. The MSE difference between the production output and the desired production is taken as the loss function here and the LTC controller model parameter weights are updated via back-propagation, while the LTC plant model weights are frozen and remain unchanged. Both controllers have an additional input (Figure 5), soymeal desired and oil recycled desired for the oil and diesel plants respectively, in addition to the main production signals that they are controlling for that is not included in the MSE loss calculation. The algorithmic controllers are trained over the same ASPEN Plus Dynamics synthetic data as the LTC plant models following a similar 80-20 training-to-testing split.

The LTC architecture for each algorithmic controller is shown in SI Figure 6. The number of hidden neurons for each LTC controller is chosen iteratively in the same fashion as the LTC node models. This results in 8 hidden neurons in the oil plant controller and 12 in the diesel plant controller, likely the result of the greater complexity in the diesel plant processes. The motor variables chosen for each controller LTC model correspond to the plant feedstocks (soybean and hexane for the oil plant controller and oil and water for the diesel plant controller).

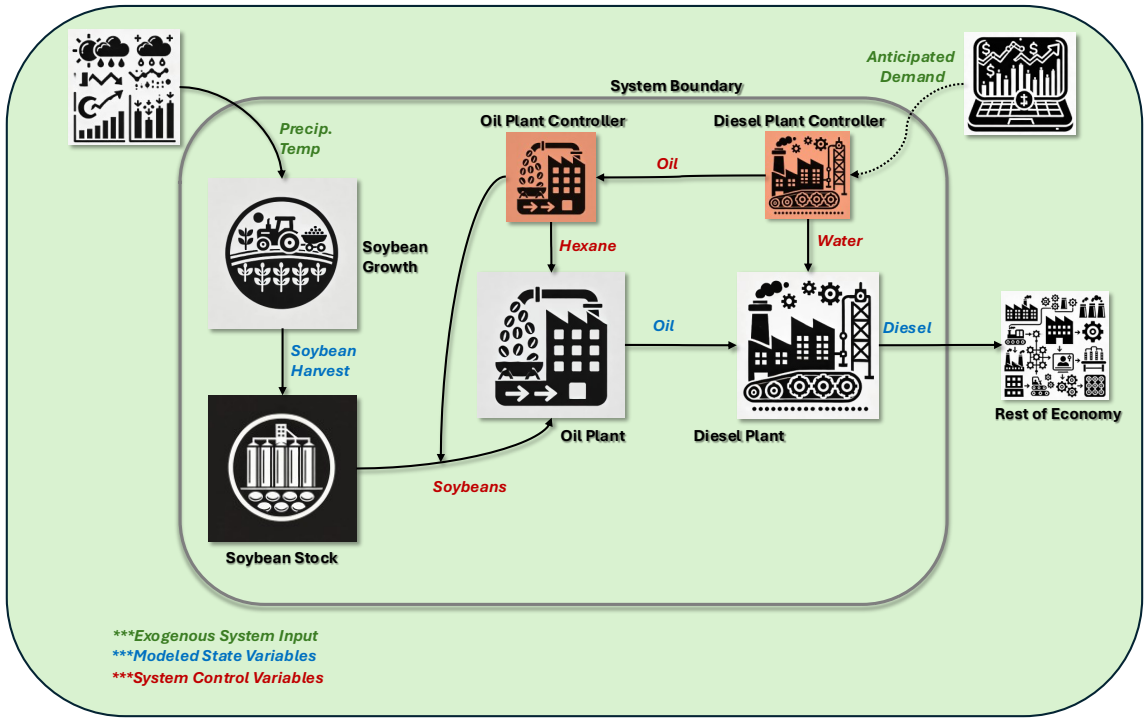

*Figure 4: Soybean network with algorithmic controllers in place to control oil and diesel production.*

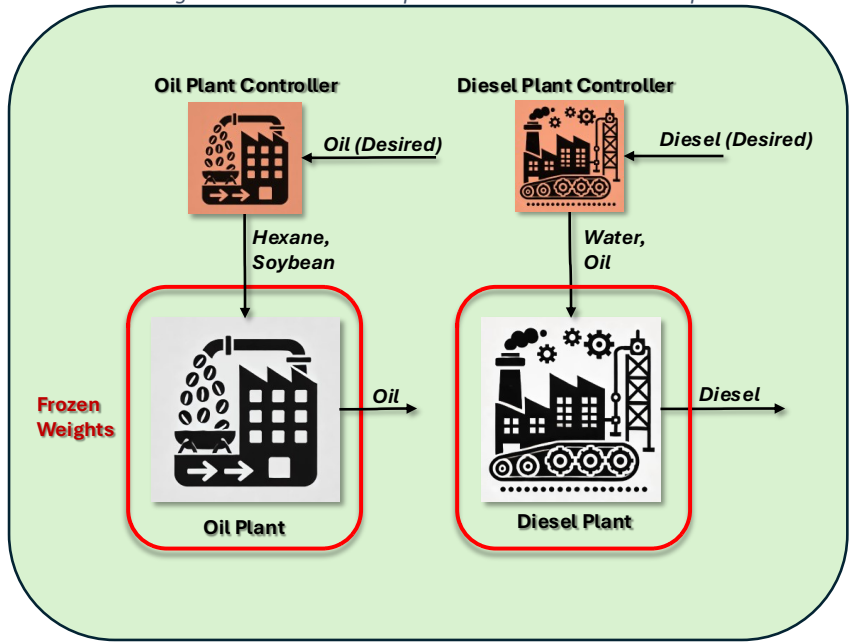

*Figure 5: Soybean oil and diesel plant controllers (LTC models) trained on top of soybean oil and diesel plant LTC models. Note: LTC model controller weights are updated via back-propagation while industrial plant LTC model weights are frozen (unchanged).*

Both the LTC models trained as oil and diesel algorithmic controllers were evaluated over the 2,000 hours corresponding to their industrial process.

The soybean oil algorithmic controller model achieved an RMSE of 0.13 on the test reference trajectory. Although it occasionally registered undershoots on outlier setpoints, it demonstrates swift reversion to optimal control values. Likewise, with an RMSE of 0.19 on the test set, the diesel algorithmic controller model echoed the robust performance observed in its counterpart, adeptly maintaining desired trajectories.

## 3. Exogenous Drivers Definition & System Simulation

**<u>Simulation & Evaluating System Resilience</u>**

Using the previously defined exogenous inputs to the system and coupled models, we evaluate the resilience of the integrated soybean production system, the dynamics across the entire life cycle stages, spanning from soybean growth to biodiesel production, under varying climate projections.

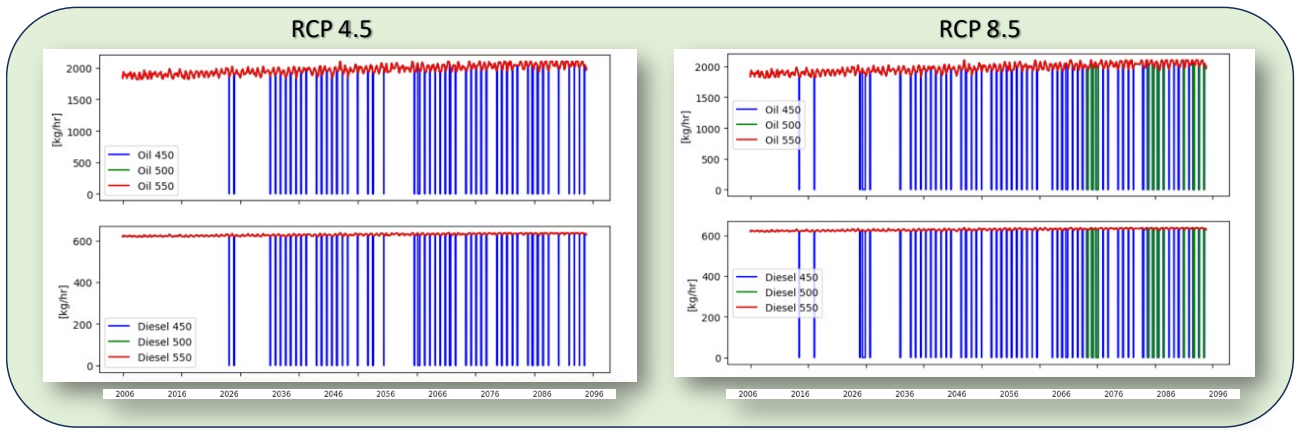

*Figure 6: Network production during RCP 4.5 and 8.5 climate scenarios with 450, 500, and 550 ha soybean farms.*

First, we focus on the network's industrial production rate over the integrated time horizon of 2006-1996. As shown in *Figure 6*, a number of differences appear between the RCP 4.5 and RCP 8.5 scenarios as well as between the 450-, 500- and 550-hectare farms. At the same farm size, we see an increased rate of failure, indicated by dips to zero, between RCP 4.5 and 8.5 as well an earlier first failure occurring around 2016 versus 2026. Further, only the 450-ha farm-supplied network exhibits failures during the RCP 4.5 scenario, whereas for RCP 8.5 both the 450 and 500 ha farm-supplied networks exhibit failure, with the 500-ha farm-supplied network beginning to fail much later around 2076.

Similarities do exist between the RCP 4.5 and 8.5 scenario production shown in *Figure 6*, with the rate of production failures increasing over time in both. Significant periods with high frequency of failure begin in both scenarios around 2030 and 2060, with denser failure rates shown in the RCP 8.5 scenario.

The cumulative waste generated between each RCP scenario also shows some similarities. In both cases, the 450-ha farm-driven network produces no waste, with the 500 and 550 each producing more respectively, as shown in *Figure 7*. However, in both the 500 and 550 ha farm-driven networks, the waste generation in the RCP 4.5 scenario is nearly double that of the RCP 8.5 by the end of the century. Additionally, in the waste accumulation plateaus earlier (around 2020) for the 500-ha farm driven RCP 8.5 scenario than the RCP 4.5 equivalent (around 2030). In the RCP 4.5 scenarios the 500-ha farm-driven network's waste never begins to plateau but continues linearly throughout the century, whereas in the RCP 8.5 scenario the waste accumulation follows a logistic slowdown to begin a plateau around 2065. Higher waste accumulation is expected in the RCP 4.5 scenario as crops are less stressed by climate variability and stock accumulation is higher, whereas in the RCP 8.5 scenario this waste is instead used during periods of lower crop production. This indicates that any capacity expansion during RCP 4.5 to use additional stock may be underutilized if RCP 8.5 scenarios materialize faster, locking in capital investments.

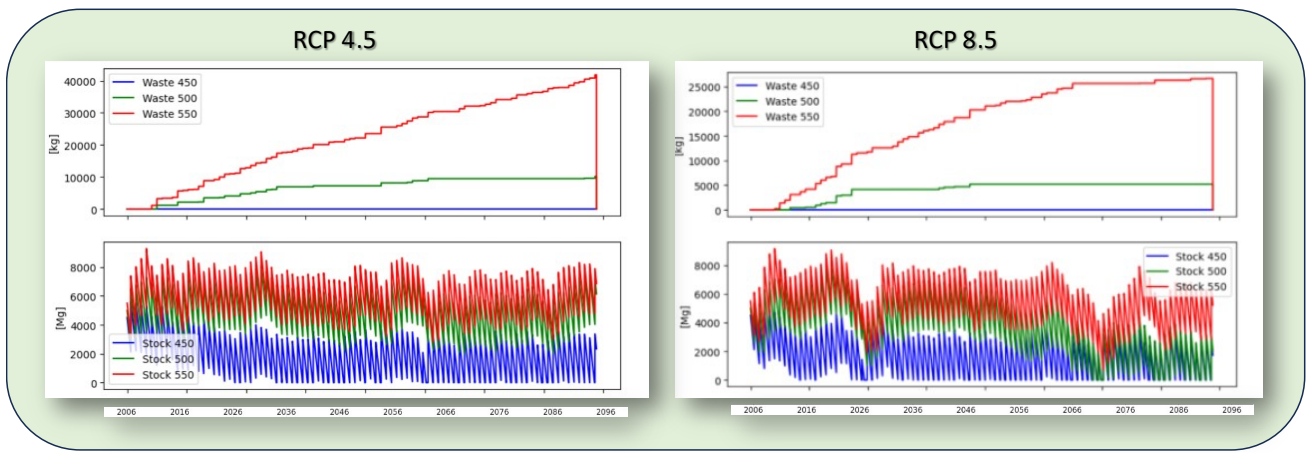

*Figure 7: Network waste and stock accumulation during RCP 4.5 and 8.5 climate scenarios with 450, 500, and 550 ha soybean farms.*

      The stock accumulation overtime, shown in *Figure 7* also reveals some key differences between scenario outcomes. While in both RCP scenarios the 450-ha farm-supplied network stock is the lowest oscillating around 2000 metric tons (Mg), in the RCP 4.5 scenarios the 500 and 550 ha farm-supplied stocks follow the same trajectory between 2006 to 2096, in the RCP 8.5 scenarios a bifurcation in stock trajectory occurs around the year 2050. After this point the 500-ha farm-supplied stock moves to follow the trajectory of the 450-ha farm-supplied stock. The bifurcation is followed by a period where the longer-term trajectory of the 550 farm-supplied network stock is also trending towards zero, before rising and following its previous pattern. A similar sharp dipping trend towards zero occurs in the RCP 8.5 scenario for the 500 and 550 ha scenarios around the year 2025, yet no bifurcation occurs before or directly after.

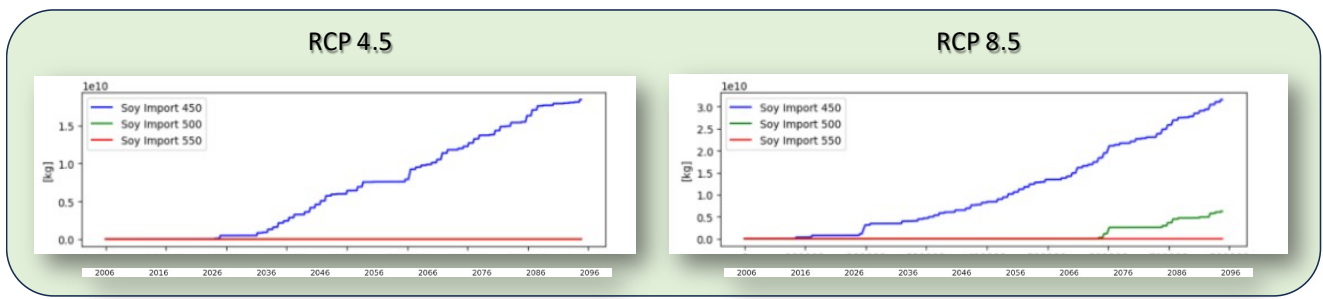

*Figure 8: Network required import such that production would continue to meet hypothetical demand during RCP 4.5 and 8.5 climate scenarios with 450, 500, and 550 ha soybean farms.*

The cumulative soybean-import required over time required for the network to remain operational (Figure 8) shows further difference between RCP scenarios. While in both scenarios the 550-ha farm-supplied network requires no exogenous import to remain operational throughout the duration of the century, the import required in the 450-ha farm driven network scenario for RCP 8.5 is nearly double that of RCP 4.5. Furthermore, while the net import required is linear in the RCP 4.5 scenario, we see a nonlinear increase over time in the RCP 8.5 import. The RCP 8.5 scenario with 500 ha farm-supply also begins to require import around 2070 (reflecting the failure trend shown in Figure 6).

## Conclusions and Discussions

In this work, we propose a novel approach to model dynamics of material flows in coupled natural-industrial ecosystems using a Liquid-Time Constant modeling approach for surrogate modeling of material flow dynamics at each node. This facilitates evaluating the resilience of these production networks to climate change scenarios, reliance on imports and quantifying the "waste" generation in these networks, that is necessary for design of sustainable coupled natural-industrial ecosystems.

The results reveal several key insights regarding the resilience and sustainability of the coupled soybean production network in Champaign, Illinois when subject to different climate change scenarios and farm sizes. Overall, the more extreme RCP 8.5 climate scenario led to earlier and more frequent production failures across all farm sizes compared to RCP 4.5. This aligns with expectations, as the hotter temperatures and altered precipitation patterns associated with RCP 8.5 are likely less optimal for soybean growth. However, the specific timing and frequency of failures varied non-linearly across scenarios and farm sizes, highlighting complex interdependencies. For instance, while the 450-ha farm exhibited no production failures under RCP 4.5, both the 450 ha and 500 ha farms failed under RCP 8.5. Yet the 500-ha farm did not fail until much later around 2076. This suggests potential tipping points in the network's dynamics, where smaller farms that were previously resilient begin to fail once climate extremes pass certain thresholds. Larger farms appear to withstand these shifts better initially, but eventually succumb.

The trends in waste accumulation and stock levels reinforce these non-linear behaviors. Under RCP 4.5, waste from the 500 ha and 550 ha farms increased steadily throughout the century. But for RCP 8.5, waste for these farms plateaued earlier around 2020 and 2065 respectively, indicating slowed production. Similarly, the 550-ha farm stock exhibited a precipitous drop around 2050 under RCP 8.5 before recovering, representing a potential bifurcation point. This alignment between changes in waste, stock, and production failures underscores the value of holistic assessment across the full coupled system.

Notably, the required soybean imports to sustain production were substantially higher under RCP 8.5 compared to RCP 4.5 for smaller farms. For the 450-ha farm, RCP 8.5 imports were double those for RCP 4.5. This reliance on external inputs to compensate for climate-induced growth constraints reveals vulnerabilities in network resilience. Larger farm-supplied networks were less dependent on imports, but still faced risk that can manifest in unexpected tipping points.

The algorithmic controllers developed for the soybean oil and diesel production plants demonstrated proficient tracking of specified production trajectories in response to the dynamic economic demand signal. The controllers adapted inputs like hexane and water flows as needed at each timestep to try to achieve the desired output rates, despite fluctuations in soybean inventory. This highlights the effectiveness of open-loop control schemes for emulating real-world adaptive behavior in balancing supply and demand. While not conferring resilience themselves, the controllers enabled continuous operation of the simulated network, linking natural growth variability with industrial processing requirements. Their integration provided a vital mechanism for propagating disruptions across the coupled system to assess impacts on overall production goals. The controller models thus served an important role as open-loop algorithmic constructs facilitating the analysis of resilience and sustainability under different climate scenarios.

*One limitation of* the soybean growth LTC models for the RCP 4.5 and 8.5 scenarios presented in this work, is that these provide insights into potential future growth patterns, but do not make definitive predictions. As discussed in the methodology, these models were trained on statistically downscaled climate projections to capture seasonal crop dynamics under hypothetical future warming scenarios. The models exhibited skill in mimicking complex growth patterns and key inflection points. Analysis of model outputs indicates earlier attainment of peak biomass under RCP 8.5 compared to 4.5, suggesting a need to adapt crop timelines. However, the growth models do not account for adaptive measures like shifting sowing dates, alternate crop varieties, or improved agronomic practices. Consequently, their projections should be interpreted as indicative of directional risks, not precise forecasts. Additional integrated modeling is needed to guide specific adaptive management strategies.

*Additionally,* the current system boundary focuses narrowly on soybean growth, oil production, and biodiesel output in the Champaign, Illinois area. This localized scope enables a tight-coupled physical modeling of the core production chain. However, it likely overlooks important interactions and vulnerabilities that could be included if the scope were to be increased spatially or dimensionally, which is done in other industrial ecology models such as multi-regional input-output models. Expanding spatially to encompass the broader supply chain for agricultural inputs like seeds, fertilizers, equipment etc. could reveal single points of failure. If a few key distributors are disrupted, farms across the region could be severely impacted. Transportation networks distributing finished biodiesel are also excluded, masking bottlenecks.

Incorporating crop rotation cycles would uncover potential soil nutrient depletion, pest build-up, and yield declines over successive soybean plantings. Demographic shifts driving urban expansion in the region may also increase pressure on croplands. And competition for water resources with other farms and industries is excluded. This could become critical as climate change alters precipitation and temperature patterns, whose regularity may have historically been taken for granted. Biodiversity impacts on pollinators and natural pest control services are also omitted. Their degradation could dramatically affect yields and require added pesticide inputs. Wider ecosystem impacts of agricultural runoff are similarly overlooked, which can be included by expanding the modeling of ecological systems.

**Acknowledgements**

This work was supported in part by US National Science Foundation through NSF GRFP under grant DGE-1842166 and NSF FMRG Eco 2229250.

**Conflict of Interest**

Authors declare no conflict of interest.

# Bibliography

[1]    G. P. J. Dijkema, M. ; Xu, S. ; Derrible, and R. Lifset, "Complexity and industrial ecology," *research.rug.nl*, vol. 19, no. 2, pp. 189–194, 2015, doi: 10.1111/jiec.12280.

[2]    S. Meerow, J. N.-J. of I. Ecology, and undefined 2015, "Resilience and complexity: A bibliometric review and prospects for industrial ecology," *Wiley Online Library*, vol. 19, no. 2, pp. 236–251, Apr. 2015, doi: 10.1111/jiec.12252.

[3]    R. San José, J. L. Pérez, R. M. González, J. Pecci, A. Garzón, and M. Palacios, "Impacts of the 4.5 and 8.5 RCP global climate scenarios on urban meteorology and air quality: Application to Madrid, Antwerp, Milan, Helsinki and London," *J Comput Appl Math*, vol. 293, pp. 192–207, Feb. 2016, doi: 10.1016/J.CAM.2015.04.024.

[4]    M. N. Anjum, Y. Ding, and D. Shangguan, "Simulation of the projected climate change impacts on the river flow regimes under CMIP5 RCP scenarios in the westerlies dominated belt, northern Pakistan," *Atmos Res*, vol. 227, pp. 233–248, Oct. 2019, doi: 10.1016/J.ATMOSRES.2019.05.017.

[5]    A. P. Nilawar and M. L. Waikar, "Impacts of climate change on streamflow and sediment concentration under RCP 4.5 and 8.5: A case study in Purna river basin, India," *Science of The Total Environment*, vol. 650, pp. 2685–2696, Feb. 2019, doi: 10.1016/J.SCITOTENV.2018.09.334.

[6]    O. N. Nasonova, Y. M. Gusev, E. E. Kovalev, and G. V. Ayzel, "Climate change impact on streamflow in Large-Scale River Basins: Projections and their uncertainties sourced from GCMs and RCP scenarios," *Proceedings of the International Association of Hydrological Sciences*, vol. 379, pp. 139–144, Jun. 2018, doi: 10.5194/PIAHS-379-139-2018.

[7]    J. Kim, J. Choi, C. Choi, and S. Park, "Impacts of changes in climate and land use/land cover under IPCC RCP scenarios on streamflow in the Hoeya River Basin, Korea," *Science of The Total Environment*, vol. 452–453, pp. 181–195, May 2013, doi: 10.1016/J.SCITOTENV.2013.02.005.

[8]    Y. Huang, Y. Ma, T. Liu, and M. Luo, "Climate Change Impacts on Extreme Flows Under IPCC RCP Scenarios in the Mountainous Kaidu Watershed, Tarim River Basin," *Sustainability 2020, Vol. 12, Page 2090*, vol. 12, no. 5, p. 2090, Mar. 2020, doi: 10.3390/SU12052090.

[9]    J. Wang, L. Hu, D. Li, and M. Ren, "Potential Impacts of Projected Climate Change under CMIP5 RCP Scenarios on Streamflow in the Wabash River Basin," *Advances in Meteorology*, vol. 2020, 2020.

[10]   Y. Zhang, Q. You, C. Chen, and J. Ge, "Impacts of climate change on streamflows under RCP scenarios: A case study in Xin River Basin, China," *Atmos Res*, vol. 178–179, pp. 521–534, Sep. 2016, doi: 10.1016/J.ATMOSRES.2016.04.018.

[11]   W. Temesgen Bekele, A. T. Haile, and T. Rientjes, "Impact of climate change on the streamflow of the Arjo-Didessa catchment under RCP scenarios," *Journal of Water and Climate Change |*, vol. 12, p. 2021, 2021, doi: 10.2166/wcc.2021.307.

[12]   W. Rosenthal, D. Ort, and E. A. Ainsworth, "Simulating Soybean-Water Relations under Elevated Atmospheric CO2: A Synthesis of Modeling and Experimental Approaches," *Plant Cell Environ*, 2019.

[13]  Z. Liu, G. Hoogenboom, K. Hu, and B. Li, "Simulation of Soybean Growth and Yield in Variable Weather," *Journal of Agricultural Science*, 2016.

[14]  M. H. Costa, T. Yanagi, M. A. Martins, and P. C. Sentelhas, "Modeling the potential productivity of soybean under climate change in Brazil," *Science of The Total Environment*, vol. 682, pp. 155–170, 2019.

[15]  H. Yang, A. Dobermann, K. G. Cassman, and D. T. Walters, "Modeling Soybean Growth and Yield under Future Climate Change and Adaptation Technologies," *Agron J*, 2019.

[16]  G. L. Hammer *et al.*, "Plant Growth and Climate Change," *J Exp Bot*, 2004.

[17]  S. Beck, M. M.-W. I. R. Climate, and undefined 2018, "The IPCC and the new map of science and politics," *Wiley Online Library*, vol. 9, no. 6, Nov. 2018, doi: 10.1002/wcc.547.

[18]  J. Rogelj, M. Meinshausen, and R. Knutti, "Global warming under old and new scenarios using IPCC climate sensitivity range estimates," *Nature Climate Change 2012 2:4*, vol. 2, no. 4, pp. 248–253, Feb. 2012, doi: 10.1038/nclimate1385.

[19]  L. Cioni, "The roles of System Dynamics in environmental problem solving," *Syst Dyn Rev*, vol. 18, no. 3, pp. 283–296, 2002.

[20]  S. Pauliuk, G. Majeau-Bettez, and D. B. Müller, "A General System Structure and Accounting Framework for Socioeconomic Metabolism," *J Ind Ecol*, vol. 19, no. 5, pp. 728–741, Oct. 2015, doi: 10.1111/JIEC.12306.

[21]  T. E. Graedel, "Material Flow Analysis from Origin to Evolution," *Environ Sci Technol*, vol. 53, no. 21, pp. 12188–12196, Nov. 2019, doi: 10.1021/ACS.EST.9B03413.

[22]  C. Sendra, X. Gabarrell, and T. Vicent, "Material flow analysis adapted to an industrial area," *J Clean Prod*, vol. 15, no. 17, pp. 1706–1715, 2007, doi: 10.1016/J.JCLEPRO.2006.08.019.

[23]  D. E. Rivera, H. Lee, M. W. Braun, and H. D. Mittelmann, "' Plant-Friendly' system identification: a challenge for the process industries," *IFAC Proceedings Volumes*, vol. 36, no. 16, pp. 891–896, 2003.

[24]  F. Harirchi *et al.*, "On sparse identification of complex dynamical systems: A study on discovering influential reactions in chemical reaction networks," *Fuel*, vol. 279, p. 118204, 2020.

[25]  D. E. Rivera, "System identification in the process industries," *Automatica*, vol. 32, no. 12, pp. 1661–1676, 1996.

[26]  C. Davis, I. Nikolíc, and G. P. J. Dijkema, "Integration of Life Cycle Assessment Into Agent-Based Modeling," *J Ind Ecol*, vol. 13, no. 2, pp. 306–325, Apr. 2009, doi: 10.1111/J.1530-9290.2009.00122.X.

[27]  C. J. Meinrenken, B. C. Sauerhaft, A. N. Garvan, and K. S. Lackner, "Combining Life Cycle Assessment with Data Science to Inform Portfolio-Level Value-Chain Engineering," *J Ind Ecol*, vol. 18, no. 5, pp. 641–651, Oct. 2014, doi: 10.1111/JIEC.12182.

[28] Z. Cao, L. Shen, S. Zhong, L. Liu, H. Kong, and Y. Sun, "A Probabilistic Dynamic Material Flow Analysis Model for Chinese Urban Housing Stock," *J Ind Ecol*, vol. 22, no. 2, pp. 377–391, Apr. 2018, doi: 10.1111/JIEC.12579.

[29] Y. Xu, Y. Geng, X. Tian, W. Shen, and Z. Gao, "Uncovering the key features of platinum metabolism in China during 2001–2022: A dynamic material flow analysis," *J Clean Prod*, vol. 446, p. 141323, Mar. 2024, doi: 10.1016/j.jclepro.2024.141323.

[30] D. B. Müller, H. P. Bader, and P. Baccini, "Long-term Coordination of Timber Production and Consumption Using a Dynamic Material and Energy Flow Analysis," *J Ind Ecol*, vol. 8, no. 3, pp. 65–88, Jul. 2004, doi: 10.1162/1088198042442342.

[31] V. S. Espinoza, S. Erbis, L. Pourzahedi, M. J. Eckelman, and J. A. Isaacs, "Material flow analysis of carbon nanotube lithium-ion batteries used in portable computers," *ACS Sustain Chem Eng*, vol. 2, no. 7, pp. 1642–1648, Jul. 2014, doi: 10.1021/SC500111Y.

[32] Y. Gao and A. C. Kokossis, "Agent-based intelligent system development for decision support in chemical process industry," in *Computer Aided Chemical Engineering*, Elsevier, 2005, pp. 1387–1392. doi: 10.1016/s1570-7946(05)80073-2.

[33] S. H. Chen, C. L. Chang, and Y. R. Du, "Agent-based economic models and econometrics," *Knowl Eng Rev*, vol. 27, no. 2, pp. 187–219, Jun. 2012, doi: 10.1017/S0269888912000136.

[34] M. R. Ghali, J. M. Frayret, and C. Ahabchane, "Agent-based model of self-organized industrial symbiosis," *J Clean Prod*, vol. 161, pp. 452–465, Sep. 2017, doi: 10.1016/J.JCLEPRO.2017.05.128.

[35] E. Bonabeau, "Agent-based modeling: Methods and techniques for simulating human systems," *Proceedings of the National Academy of Sciences*, vol. 99, no. suppl_3, pp. 7280–7287, May 2002, doi: 10.1073/PNAS.082080899.

[36] T. Filatova, P. H. Verburg, D. C. Parker, and C. A. Stannard, "Spatial agent-based models for socio-ecological systems: Challenges and prospects," *Environmental Modelling & Software*, vol. 45, pp. 1–7, Jul. 2013, doi: 10.1016/J.ENVSOFT.2013.03.017.

[37] K. P. H. Lange, G. Korevaar, I. Nikolic, and P. M. Herder, "Actor behaviour and robustness of industrial symbiosis networks: An agent-based modelling approach," *JASSS*, vol. 24, no. 3, p. 8, Jun. 2021, doi: 10.18564/jasss.4635.

[38] V. Albino, L. Fraccascia, and I. Giannoccaro, "Exploring the role of contracts to support the emergence of self-organized industrial symbiosis networks: An agent-based simulation study," *J Clean Prod*, vol. 112, pp. 4353–4366, 2016, doi: 10.1016/J.JCLEPRO.2015.06.070.

[39] K. J. Boote and L. H. Allen Jr, "Cropping Strategies for Efficient Use of Water and Nitrogen," *Agronomy Monographs*, 1983.

[40] K. J. Boote, J. W. Jones, and N. B. Pickering, "Cropping system models: a review of present status and future prospects," *Agron J*, vol. 88, no. 5, pp. 697–713, 1996.

[41] D. Chen, Y. Huang, Z. Li, L. Duan, and L. Ma, "Big data challenges and opportunities in crop modelling," *J Exp Bot*, vol. 65, no. 6, pp. 1737–1750, 2014.

[42] G. L. Hammer, D. R. Jordan, and J. W. Mjelde, "Advances in simulation of plant growth and crop management," *Agron J*, vol. 94, no. 1, pp. 3–36, 2002.

[43] B. D. Wardlow, S. L. Egbert, J. H. Kastens, and A. Aragon, "Machine learning for crop yield prediction in the US Corn Belt," *Environmental Research Letters*, vol. 13, no. 10, p. 104004, 2018.

[44] R. Hasani, M. Lechner, A. Amini, D. Rus, and R. Grosu, "Liquid Time-constant Networks," *Proceedings of the AAAI Conference on Artificial Intelligence*, vol. 35, no. 9, pp. 7657–7666, May 2021, doi: 10.1609/AAAI.V35I9.16936.

[45] R. Hasani, M. Lechner, A. Amini, D. Rus, and R. Grosu, "A natural lottery ticket winner: Reinforcement learning with ordinary neural circuits," *proceedings.mlr.press*, 2020, Accessed: Nov. 09, 2023. [Online]. Available: https://proceedings.mlr.press/v119/hasani20a.html

[46] M. Lechner, R. Hasani, A. Amini, T. A. Henzinger, D. Rus, and R. Grosu, "Neural circuit policies enabling auditable autonomy," *nature.com*, 2020, doi: 10.1038/s42256-020-00237-3.

[47] M. Bidollahkhani, F. Atasoy, and H. Abdellatef, "LTC-SE: Expanding the Potential of Liquid Time-Constant Neural Networks for Scalable AI and Embedded Systems," Apr. 2023, Accessed: Nov. 05, 2023. [Online]. Available: http://arxiv.org/abs/2304.08691

[48] R. Hasani *et al.*, "Closed-form continuous-time neural networks," *Nat Mach Intell*, vol. 4, no. 11, 2022, doi: 10.1038/s42256-022-00556-7.

[49] K. Riahi *et al.*, "RCP 8.5—a scenario of comparatively high greenhouse gas emissions," *Clim Change*, vol. 109, no. 1–2, pp. 33–57, 2011.

[50] E. Nabavi, K. A. Daniell, and H. Najafi, "Boundary matters: the potential of system dynamics to support sustainability?," *Sustain Sci*, vol. 13, no. 1, pp. 47–63, 2018.

[51] H. Salehinejad, S. Sankar, J. Barfett, E. Colak, and S. Valaee, "Recent Advances in Recurrent Neural Networks," Dec. 2017, Accessed: Nov. 13, 2023. [Online]. Available: https://arxiv.org/abs/1801.01078v3

[52] B. Lindemann, T. Müller, H. Vietz, N. Jazdi, and M. Weyrich, "A survey on long short-term memory networks for time series prediction," *Procedia CIRP*, vol. 99, pp. 650–655, Jan. 2021, doi: 10.1016/J.PROCIR.2021.03.088.

[53] D. P. Searson, D. E. Leahy, and M. J. Willis, "GPTIPS: an open source genetic programming toolbox for multigene symbolic regression," in *Proceedings of the International multiconference of engineers and computer scientists*, 2010, pp. 77–80.

[54] S. L. Brunton, J. L. Proctor, and J. N. Kutz, "Discovering governing equations from data by sparse identification of nonlinear dynamical systems," *Proceedings of the national academy of sciences*, vol. 113, no. 15, pp. 3932–3937, 2016.

[55] W. Farlessyost and S. Singh, "Reduced order dynamical models for complex dynamics in manufacturing and natural systems using machine learning," *Nonlinear Dyn*, vol. 110, no. 2, pp. 1613–1631, Oct. 2022, doi: 10.1007/S11071-022-07695-X/FIGURES/11.

[56] R. Subramanian, R. R. Moar, and S. Singh, "White-box Machine learning approaches to identify governing equations for overall dynamics of manufacturing systems: A case study on distillation column," *Machine Learning with Applications*, vol. 3, p. 100014, Mar. 2021, doi: 10.1016/J.MLWA.2020.100014.

[57] R. P. Anex, C. J. Kucharik, M. L. Ruffo, J. D. Muñoz, and R. D. Jackson, "BioCro: A simulation model of crop productivity for the US Midwest," *Agric Syst*, vol. 141, pp. 96–105, 2015.

[58] M. L. Matthews, A. Marshall-Colón, J. M. McGrath, E. B. Lochocki, and S. P. Long, "Soybean-BioCro: a semi-mechanistic model of soybean growth," *In Silico Plants*, vol. 4, no. 1, pp. 1–11, Jan. 2022, doi: 10.1093/INSILICOPLANTS/DIAB032.

[59] J. T. Abatzoglou and T. J. Brown, "A comparison of statistical downscaling methods suited for wildfire applications," *International journal of climatology*, vol. 32, no. 5, pp. 772–780, 2012.

[60] F. W. Chen and C. W. Liu, "Estimation of the spatial rainfall distribution using inverse distance weighting (IDW) in the middle of Taiwan," *Paddy and Water Environment*, vol. 10, no. 3, pp. 209–222, Sep. 2012, doi: 10.1007/S10333-012-0319-1/FIGURES/7.

[61] J. T. Abatzoglou and T. J. Brown, "A comparison of statistical downscaling methods suited for wildfire applications," *International Journal of Climatology*, vol. 32, no. 5, pp. 772–780, Apr. 2012, doi: 10.1002/JOC.2312/FULL.